\documentclass[aps,preprint,groupedaddress,showpacs]{revtex4-1}
\pdfoutput=1

\usepackage{graphicx}
\usepackage{bm}

\usepackage{hyperref}

\newcommand{\fscale}{0.6}
\newcommand{\gscale}{1.425}
\newcommand{\hscale}{0.9}
\newcommand{\iscale}{0.5}
\newcommand{\kscale}{\iscale}

\begin{document}

\title{A Magnetization Sensitive Potential at Garnet-Metal Interfaces}

\author{L. R. Hunter}
\affiliation{Department of Physics, Amherst College, Amherst, MA 01002}
\author{K. A. Virgien}
\affiliation{Department of Physics, Amherst College, Amherst, MA 01002}
\author{A. W. Bridges}
\affiliation{Department of Physics, Amherst College, Amherst, MA 01002}
\author{B. J. Heidenreich}
\thanks{Present address: Department of Physics, Cornell University}
\affiliation{Department of Physics, Amherst College, Amherst, MA 01002}
\author{J. E. Gordon}
\affiliation{Department of Physics, Amherst College, Amherst, MA 01002}
\author{A. O. Sushkov}
\affiliation{Yale University, Department of Physics, P.O. Box 208120, New Haven, CT 06520-8120}

\date{\today}

\begin{abstract}
We investigate a magnetization-dependent voltage that appears at the interface between garnets and various metals. The voltage is even in the applied magnetic field and is dependent on the surface roughness and the pressure holding the surfaces together. Large variations in the size, sign and magnetic dependence are observed between different metal surfaces. Some patterns have been identified in the measured voltages and a simple model is described that can accommodate the gross features. The bulk magnetoelectric response of one of our polycrystalline YIG samples is measured and is found to be consistent with a term in the free energy that is quadratic in both the electric and magnetic fields. However, the presence of such a term does not fully explain the complex magnetization dependence of the measured voltages.
\end{abstract}



\maketitle
\section{Introduction}

Surface magnetoelectric effects are of considerable interest and have been the subject of a number of recent investigations.\cite{PhysRevLett.101.137201,PhysRevB.79.140403} Enhanced magnetoelectric couplings have been observed in bilayers of yttrium iron garnet (YIG) and lead magnesium niobate-lead titanate.\cite{srinivasan:222506} Large magnetoelectric couplings have been studied in new multiferroic materials.\cite{choong:2007ab,ramesh:2007ab} Much of the renewed interest in the magneto-electric effect has been driven by the possibility of using this effect for controlling magnetic data storage with an electric field.\cite{zhao:2006ab,Zavaliche:2007fj}

An effort to measure the electron electric dipole moment (edm) using gadolinium doped YIG is underway.\cite{PhysRevLett.95.253004} Presently, the experimental precision is limited by an unanticipated electrical potential, dubbed the ``$M$-even effect'', that is predominantly symmetric in $H$, the applied magnetic field. With the hope of overcoming this obstacle, we have launched an investigation into the characteristics of this potential. The effect is observed to be highly sensitive to the roughness of the garnet surfaces and the pressure applied to hold these surfaces together. A complex magnetic dependence is observed that varies dramatically with the composition of the metallic electrode or epoxy in immediate contact with the garnet surface. Overwhelmingly, the evidence suggests that the effect is a surface magnetoelectric effect. 

\section{Observations in G\lowercase{d}IG Toroids}

\begin{figure}
\includegraphics[scale=\fscale]{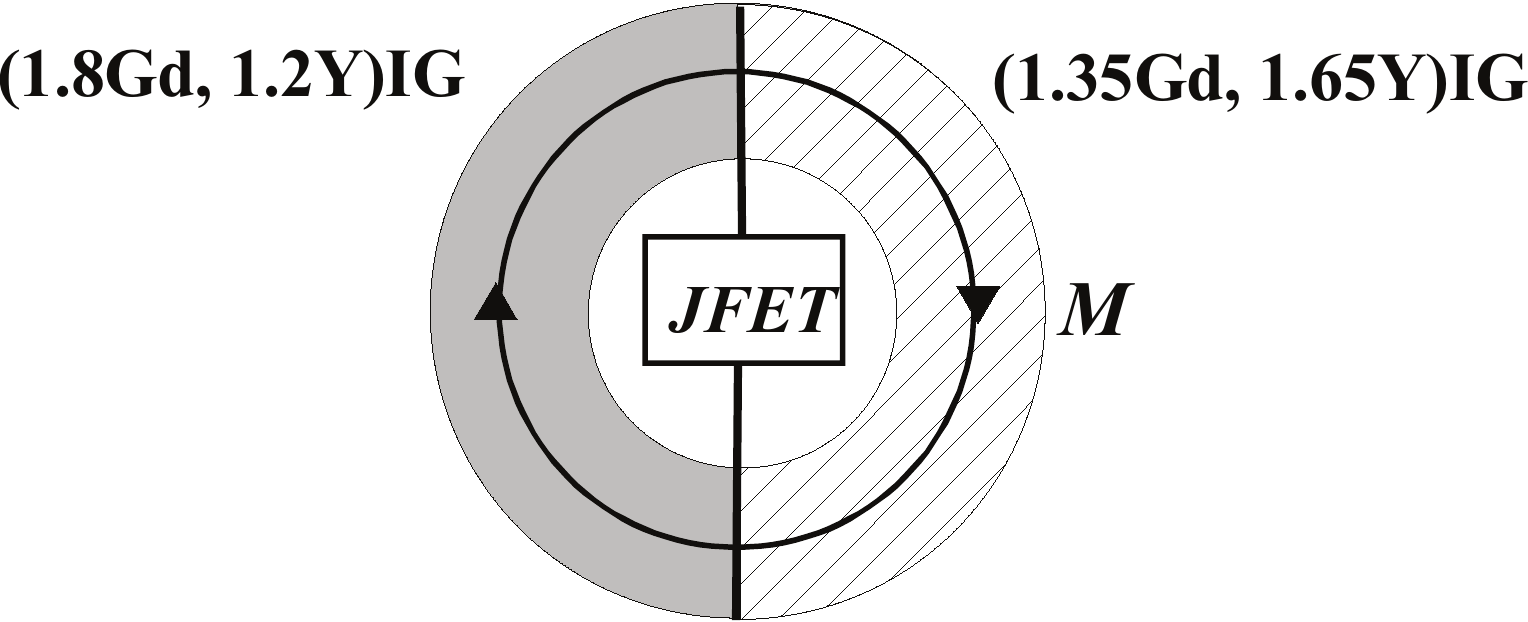}
\caption{The GdIG toroid. The applied field and magnetization circulate around the toroid.\label{fig1}}
\end{figure}

We first observed the $M$-even effect in the magnetic toroid designed to search for the electron edm (Fig.~\ref{fig1}). The sample was assembled from two ``C''s of gadolinium-yttrium-iron-garnet (Gd$_x\,$Y$_{(3-x)}\,$Fe$_2\,$Fe$_3\,$O$_{12}$) that were geometrically identical. ``C1'' had $x = 1.35$ while ``C2'' had $x = 1.8$ where $x$ represents the average number of Gd ions per formula unit. Copper foil electrodes were bonded between the two ``C''s using silver epoxy. A coil wrapped on the outside of a copper Faraday cage was used to produce $H$ in the toroidal direction. The voltage on each electrode was monitored using a detector built with a Cascode pair of BF 245 JFETs. The detector has an input impedance of $10^{13}\ \Omega$ and an input capacitance of about 4 pF.

\begin{figure}
\includegraphics[scale= \kscale]{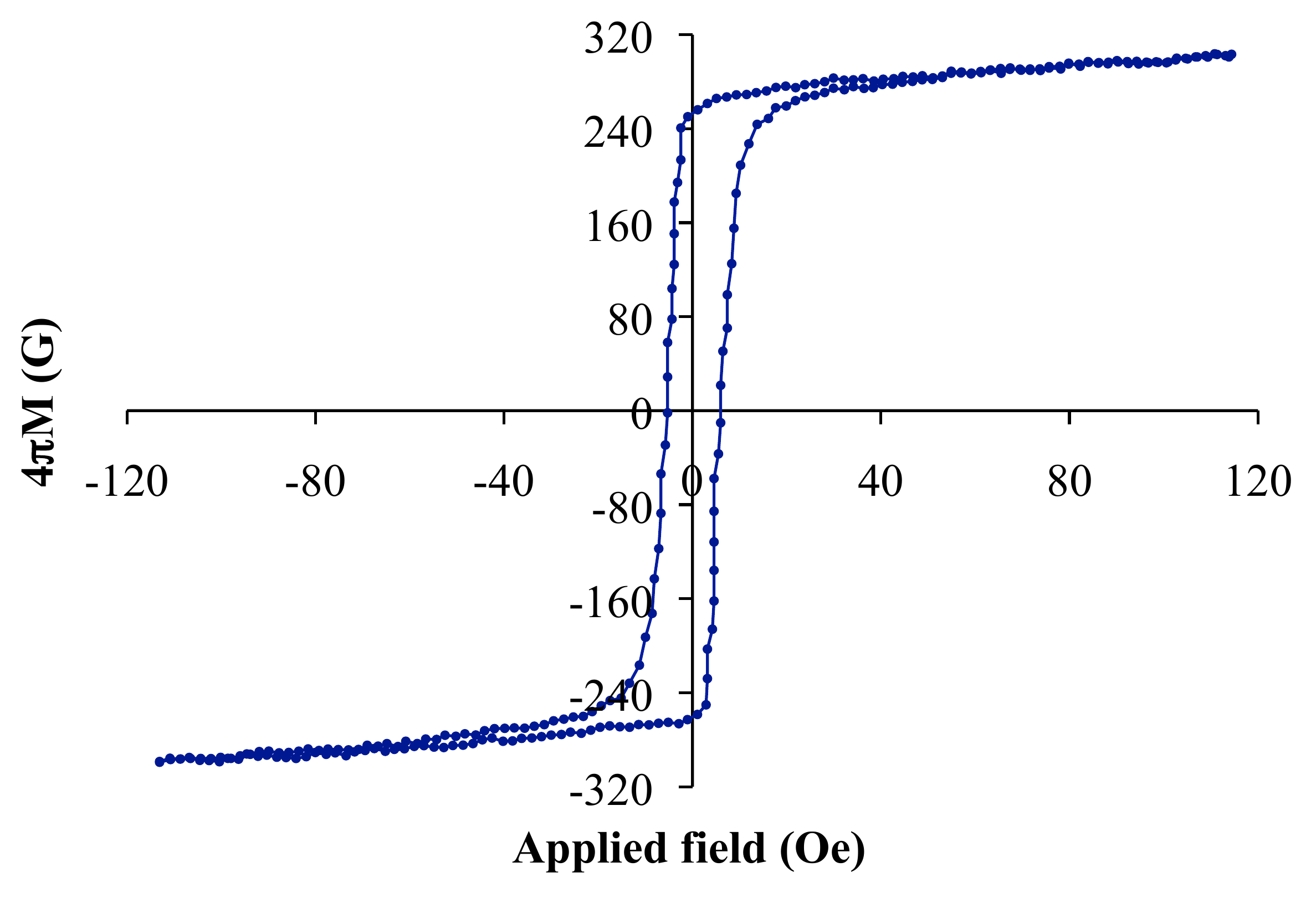}
\caption{A magnetic hysteresis curve obtained from the GdIG toroid at 127 K.\label{fig2}}
\end{figure}

A hysteresis curve inferred from coils wrapped on the toroid is shown in Figure \ref{fig2}. It is important to keep this curve in mind when viewing the subsequent data, which are generally plotted as a function of the applied magnetic field. In the toroid, the magnetization is nearly saturated at applied fields of only a few tens of Oersted.

\begin{figure}
\includegraphics[scale=\iscale]{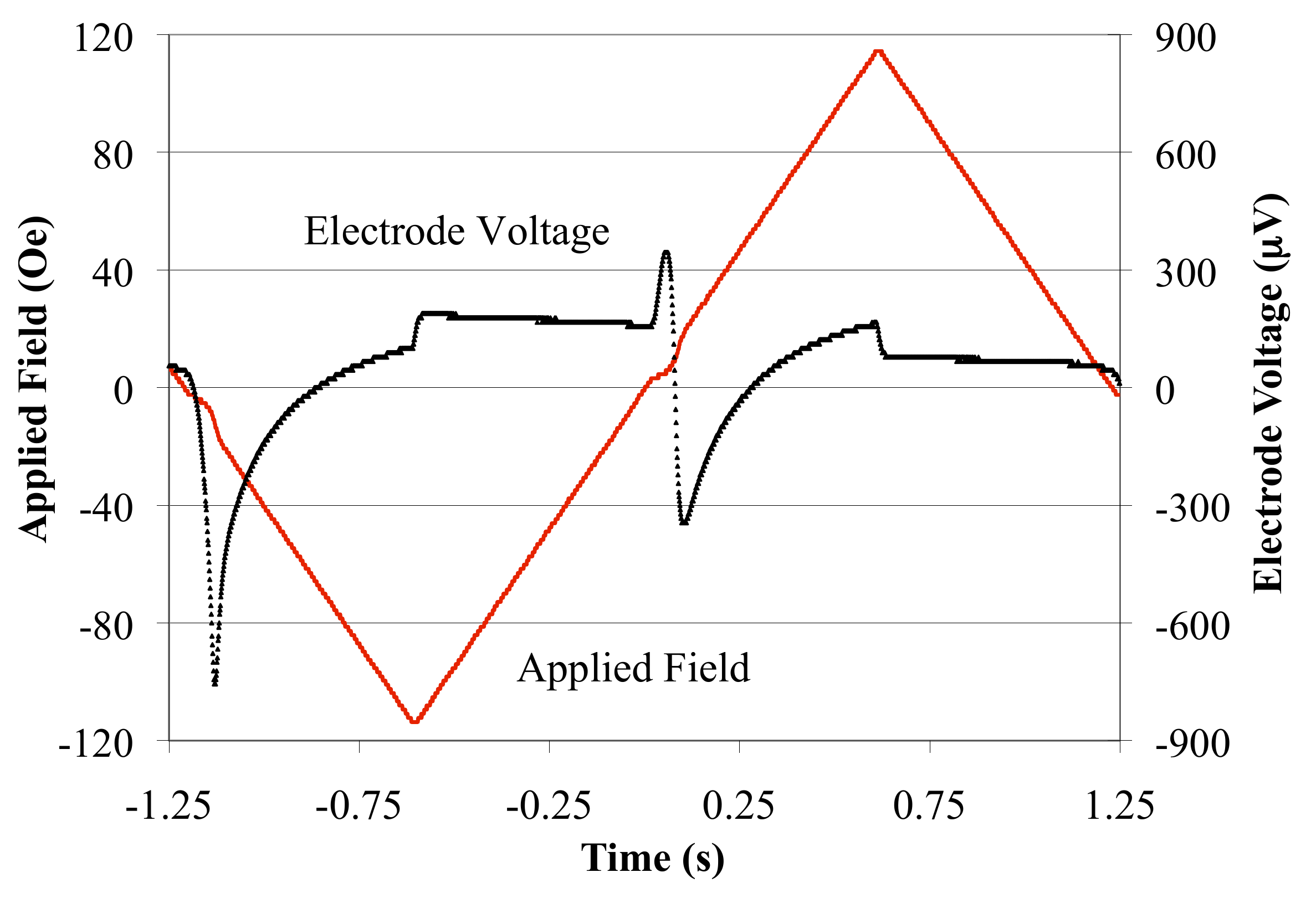}
\caption{The observed electrode voltage and applied magnetic field as a function of time. These data are taken at 127 K with an 0.41 Hz triangle wave.\label{fig3}}
\end{figure}

To investigate the $M$-even effect we monitor the electrode voltages and the sample magnetization as a function of the applied field $H$, which is varied using a triangular wave. A typical trace is shown in Fig.~\ref{fig3}. The bumps in the triangle wave at approximately $t = -1.2\ \mathrm{s}$ and $t = 0\ \mathrm{s}$ are due to the back emf associated with the reversal of the YIG magnetization. 

We find that such traces are best understood by splitting them into ``even'' and ``odd'' components. Denoting the electrode voltage by a function $V(t)$ with period $T$, we define the even and odd components of $V(t)$ to be
\begin{eqnarray*}
V_{\mathrm{EVEN}}(t) & = & \frac{1}{2} \left(V(t)+V(t+T/2)\right) \\
V_{\mathrm{ODD}}(t) & = & \frac{1}{2} \left(V(t)-V(t+T/2)\right)
\end{eqnarray*}

\begin{figure}
\includegraphics[scale=\fscale]{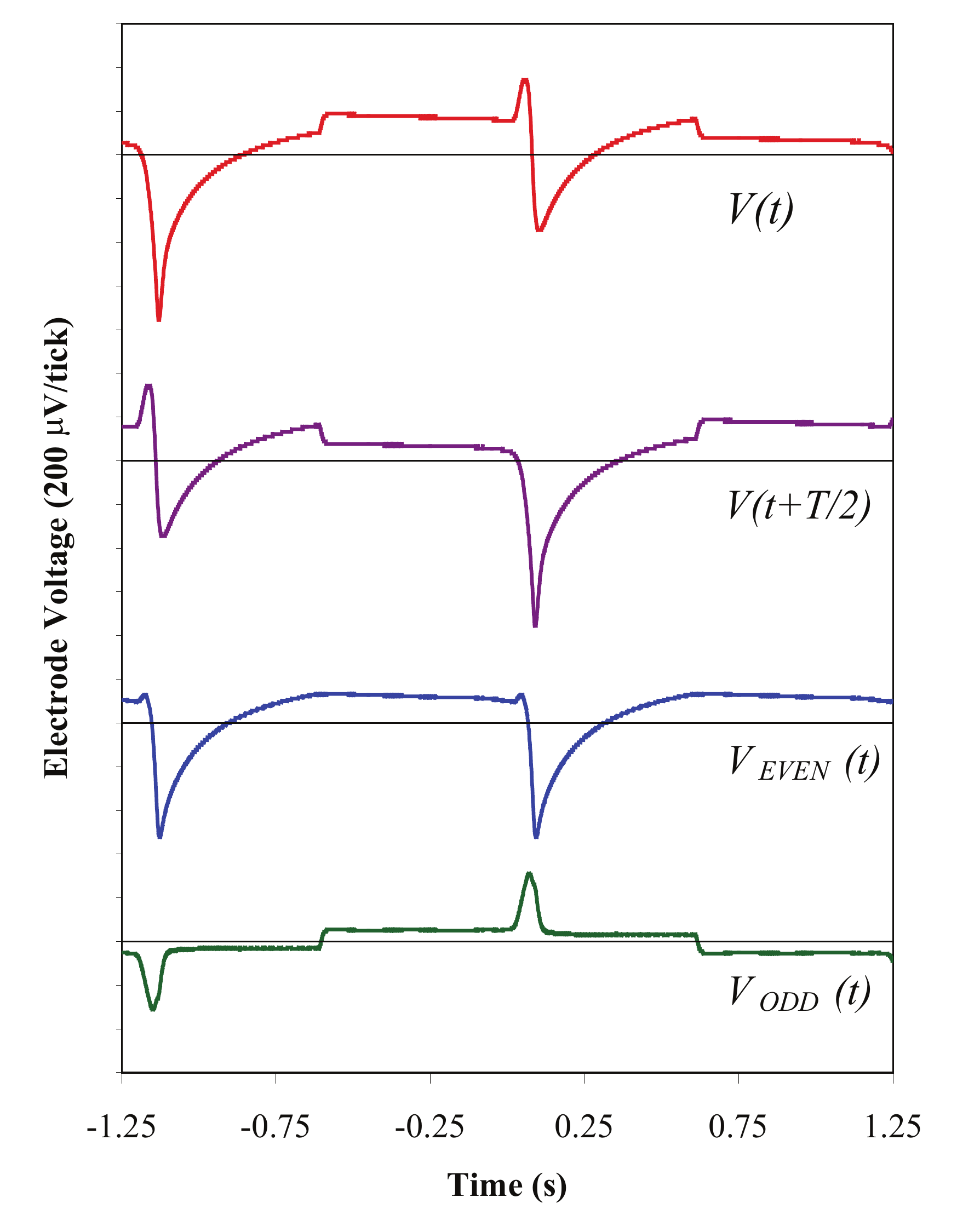}
\caption{An example of how the even and odd components of the electrode voltage are extracted. The first trace is the same as that shown in Fig.~\ref{fig3}, whereas the second is offset by half a period. The third trace (the even component) is obtained by averaging the first two traces, and the fourth (the odd component) is obtained by taking one half of their difference. The first trace can be recovered by adding the third and fourth together.\label{fig4}}
\end{figure}

It is clear that $V(t)$ is completely specified over its entire period $T$ by specifying $V_{\mathrm{EVEN}}(t)$ and $V_{\mathrm{ODD}}(t)$ over a half-period $T/2$. The process of computing the even and odd components of the electrode voltage is illustrated graphically in Figure \ref{fig4}.

\begin{figure}
\includegraphics[scale=\gscale]{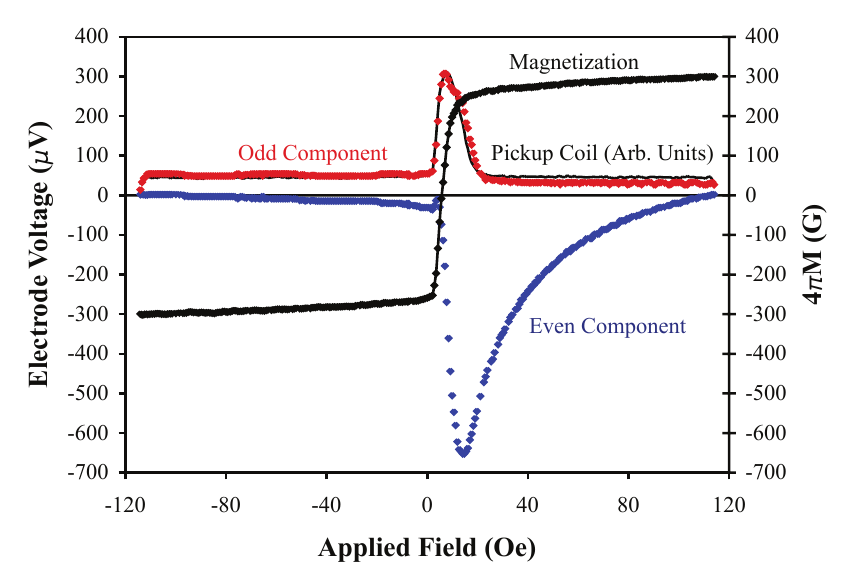}
\caption{The even and odd components of the electrode voltage shown in Fig.~\ref{fig3}, plotted over a half-period as a function of applied field. These data are acquired as the applied field is increased (left to right), and so, because of magnetic hysteresis, the even component is not symmetric about $H = 0$. Also plotted are the voltage measured by a pickup coil, rescaled to match the odd component of the electrode voltage, and the magnetization of the sample, as inferred from integrating this voltage and correcting for the effects of the driving coil.\label{fig5}}
\end{figure}

The primary benefit of analyzing the data in this fashion is that the current in the driving coil has a vanishing even component, and consequently inductive voltages on the electrodes should show up in the odd component only. Figure \ref{fig5} shows the even and odd components of the trace analyzed above plotted as a function of the applied field, as well as the voltage measured by a pickup coil wound around the toroidal Faraday cage enclosing the sample, and the sample magnetization inferred from integrating the pickup coil voltage.

Transient induction pulses due to $dB/dt$ appear clearly in the odd component of the electrode voltage. They quickly approach a constant value associated with $dH/dt$ after the reversal of the magnetic domains. As Figure \ref{fig5} demonstrates, this behavior also appears in the voltage measured across the pickup coil, suggesting that the odd component is primarily or entirely inductive in nature. As one expects, the magnitude of the odd component grows linearly with increasing scan frequency. 

The even component behaves differently, changing rapidly with the reversal of the sample magnetization and then falling off roughly as $1/H$ at higher fields. We refer to this voltage as the ``$M$-even effect'' since it appears in the even component of the voltage and is clearly associated with the reversal of the sample magnetization. In our toroidal geometry the $M$-even effect is nearly identical for the two electrodes. Unlike the odd component of the signal, the $M$-even voltage is not appreciably modified by changing the frequency of the triangle wave. The persistence of the $M$-even voltage with the increasing applied field is particularly problematic for the edm measurement. Though the effect appears to be strongly dependent on the sample magnetization, we find it more illuminating to plot the voltage as a function of the applied magnetic field as this more clearly displays the problematic $1/H$ dependence.

\begin{figure}
\includegraphics[scale=\fscale]{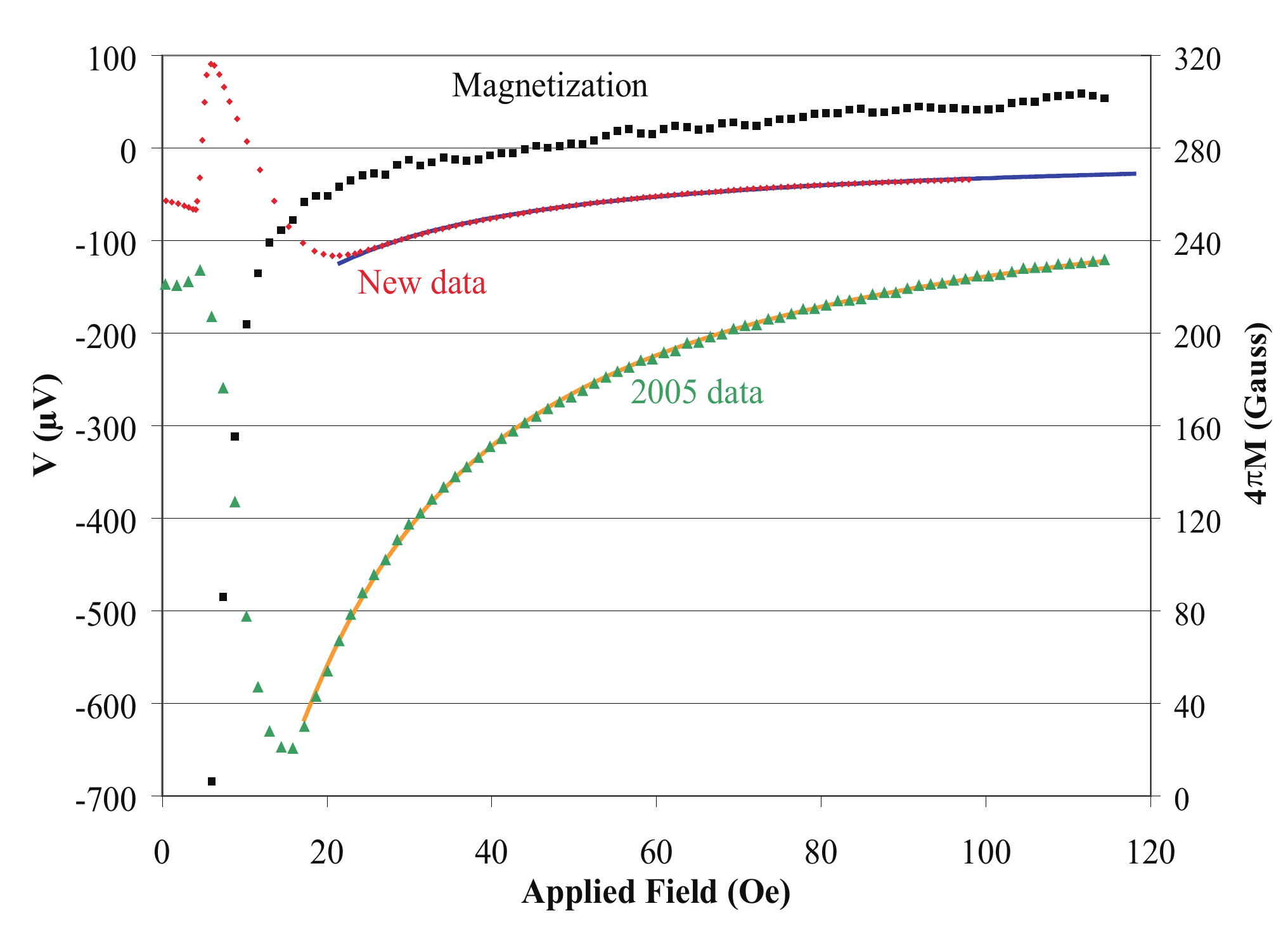}
\caption{The $M$-even voltage ($V$) observed on our GdIG sample in our 2005 data and after reassembling the toroid (new). Only the half of the cycle with increasing magnitude of $H$ is shown. The half of the cycle with decreasing magnitude of $H$ is essentially flat. The solid lines are the asymptotic fit of the high-field part of the electrode voltages to the function $V=a/H+b/H^2$. The magnetization of the sample as a function of $H$ is also shown on the secondary axis. The $M$-even voltage has an onset that coincides with the rising of the magnetization. These data were taken at 127 K.\label{fig6}}
\end{figure}

After the submission of ref.~\cite{PhysRevLett.95.253004} we disassembled our sample and repaired several flaws. A small chip that had been present on C2 was removed by grinding down the top surfaces. This reduced the toroid height by 3/16''. The epoxy was baked off, the electrodes were removed and the interstitial garnet surfaces were ground flat. New electrodes with more secure connections were then bonded to the sample with a new silver epoxy. Following the reassembly we again measured the $M$-even effect. Somewhat surprisingly, the reassembled sample exhibited an $M$-even effect that was smaller by about a factor of 5 (Fig.~\ref{fig6}). This suggested that the effect was likely not a bulk effect but was rather associated with the surface between the garnet and the electrode.

\section{Experimental Study of the $M$-even effect on YIG cylinders}

\begin{figure}
\includegraphics[scale=\hscale]{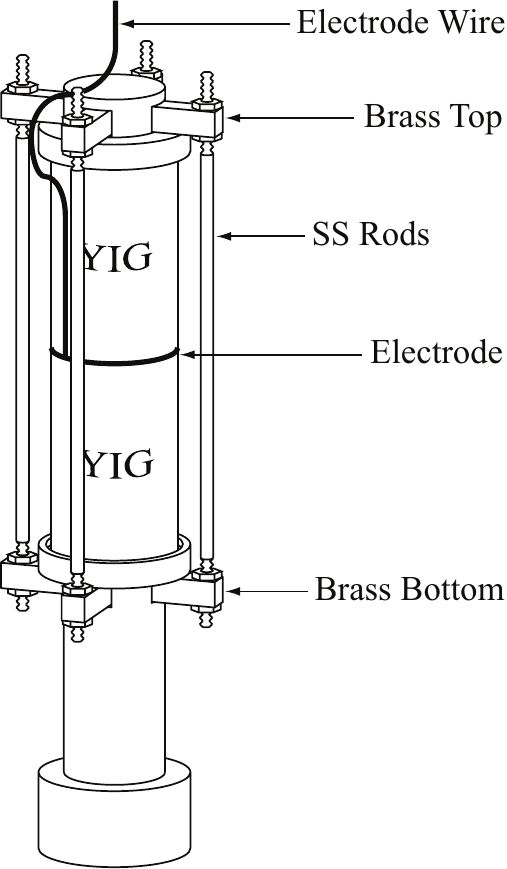}
\caption{Schematic of the assembly used to study the $M$-even effect. The stainless steel (SS) rods and the brass caps are connected to circuit ground. This entire assembly is easily mounted within a magnetic solenoid. The bore of the solenoid forms a complete Faraday cage once the ends are closed off with metal plates.\label{fig7}}
\end{figure}

Following this observation, we launched an effort to characterize the effect. First, we established that the $M$-even effect is present at room temperature in an electrode sandwiched between two cylindrical samples (1.25'' diameter, 2'' long) of YIG (Fig.~\ref{fig7}). This is important in that these cylindrical samples can be easily inserted into a solenoid magnet to study their behavior. Hence, it is not necessary to wrap a new coil for every test. In addition, operation at room temperature eliminates the time-consuming process of cycling the temperature between tests. These advantages have allowed us to study the effect in a wide variety of electrode configurations and to replace our cascode-pair detector with an instrumentation amplifier (Burr-Brown INA116). 

\begin{figure}
\includegraphics[scale=\kscale]{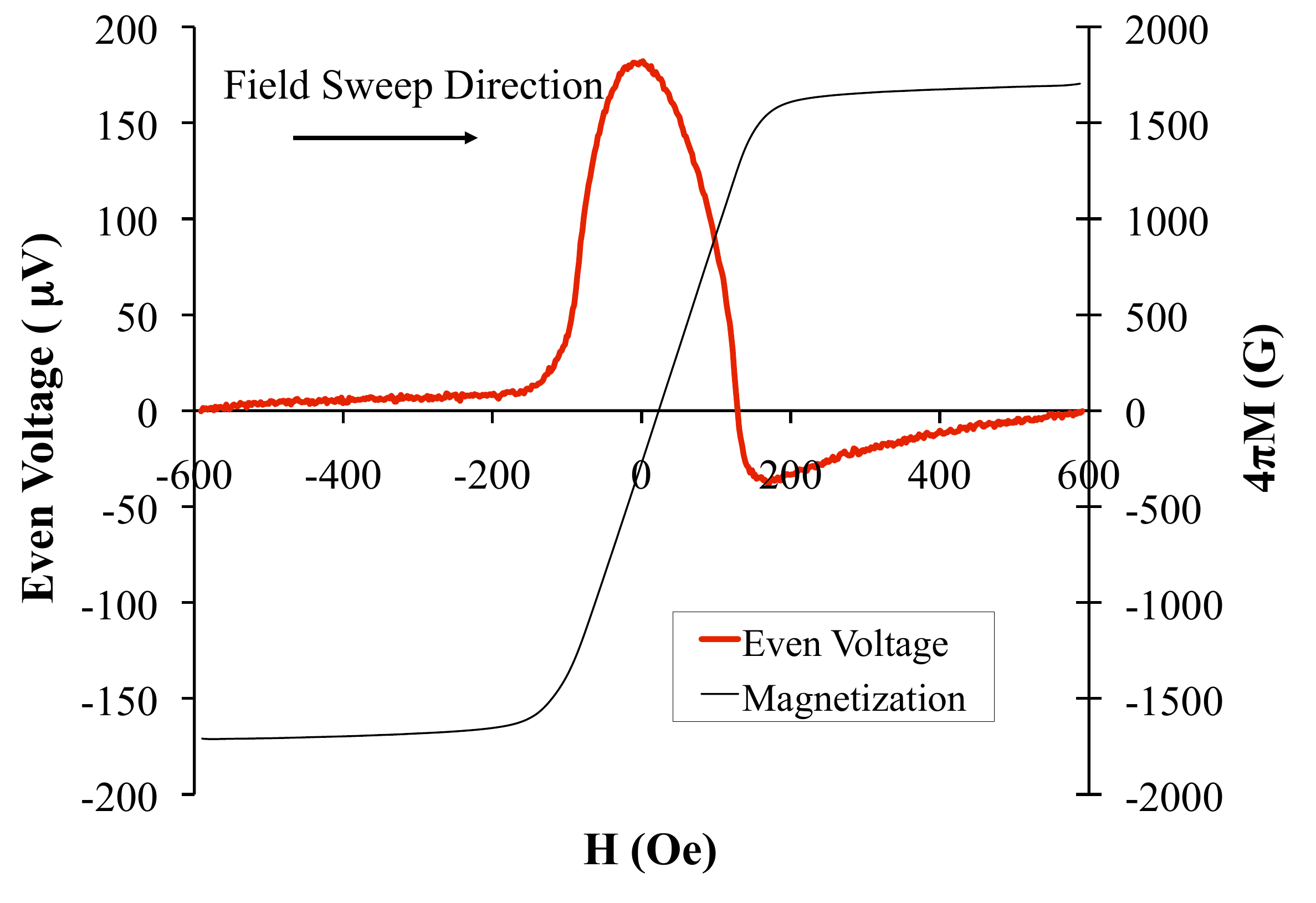}
\caption{A plot of the observed $M$-even voltage and magnetization in the new apparatus with a copper electrode between YIG cylinders. The magnitude of the applied field is increasing (decreasing) for positive (negative) values of $H$.\label{fig8}}
\end{figure}

A typical trace of the $M$-even voltage taken in this new geometry is shown in Fig.~\ref{fig8} along with a plot of the sample magnetization. In this figure and in most of the reported traces the data are an average of several independent and reproducible measurements taken on different days. The trace is essentially the signal you would observe if you scanned the applied magnetic field from -600 Oe up to 600 Oe with the odd part of the signal removed. Note that the shape of the $M$-even potential is different from that observed in the toroid. This is primarily due to the demagnetizing fields that are present in the solenoid but absent in the toroid. In the cylindrical geometry, the magnetic domains begin to relax as the magnitude of magnetic field is reduced whereas in the toroidal geometry some coercive field is required in the opposing direction before significant reduction and then reversal of the magnetization is observed. For this reason, we now display the entire plot rather than just that for increasing field as was done in Figure \ref{fig6}.

\begin{figure}
\includegraphics[scale=\kscale]{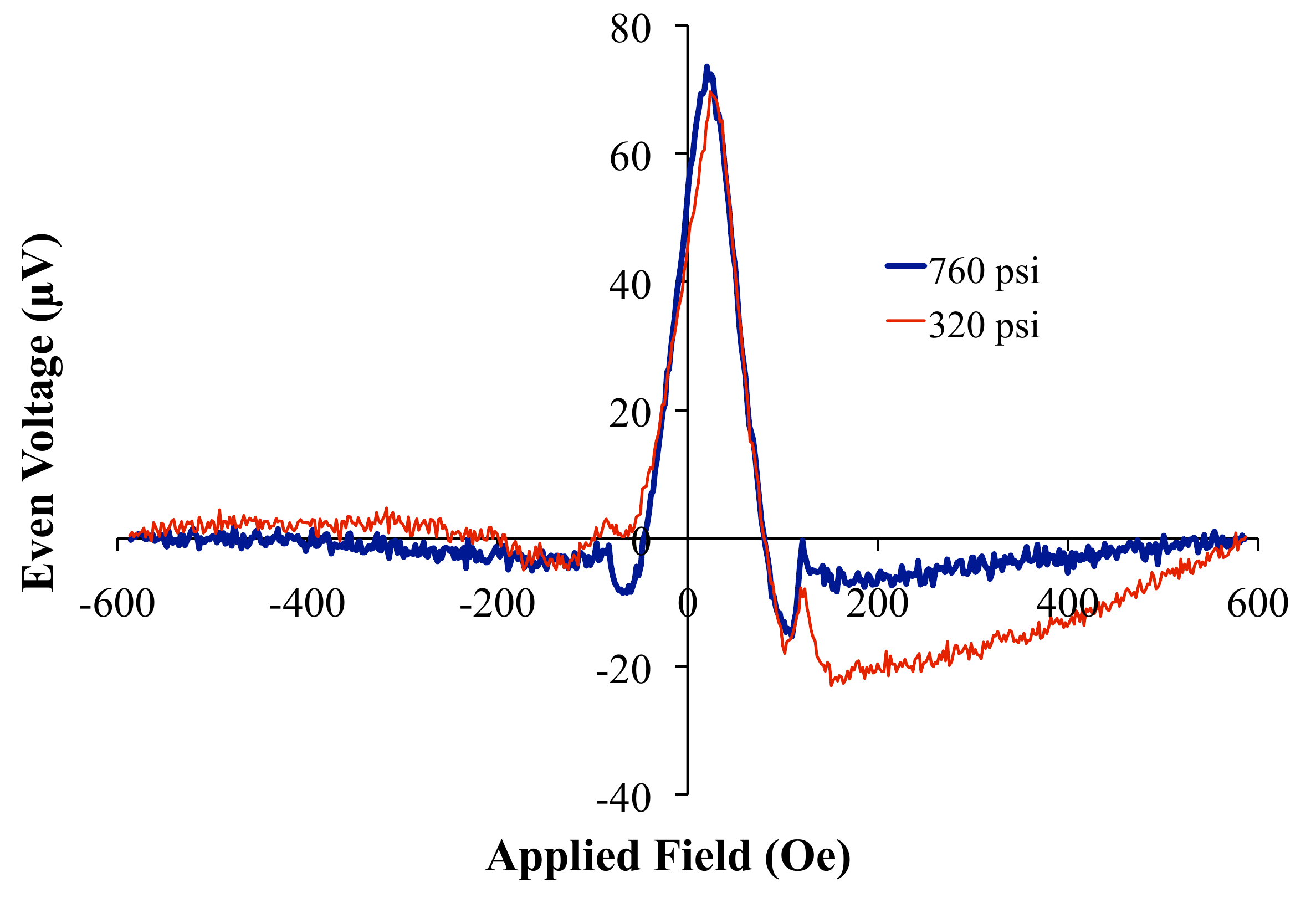}
\caption{The $M$-even voltage observed with a nickel electrode at 320 psi and 760 psi. The voltage at high, positive (increasing) magnetic fields is significantly reduced at the higher pressure.\label{fig9}}
\end{figure}

We first studied the pressure-dependence of the $M$-even effect for metallic electrodes placed between the two YIG cylinders. The pressure is varied by adjusting the tension on the 4 stainless-steel rods that hold the assembly together (Fig.~\ref{fig7}). Reproducible pressures can be achieved by plucking these rods and tuning their response to a particular frequency. We have determined the pressure that corresponds to each frequency both empirically and by calculation. All samples have been observed at multiple pressures. We find that for modest pressures (less than 300 psi), the magnitude of the $M$-even voltage generally decreases as the pressure increases. However, we observe only modest change in the $M$-even effect at higher pressures (300 psi -- 800 psi). A typical variation in this pressure regime is shown in Fig.~\ref{fig9}. 

\begin{figure}
\includegraphics[scale=\kscale]{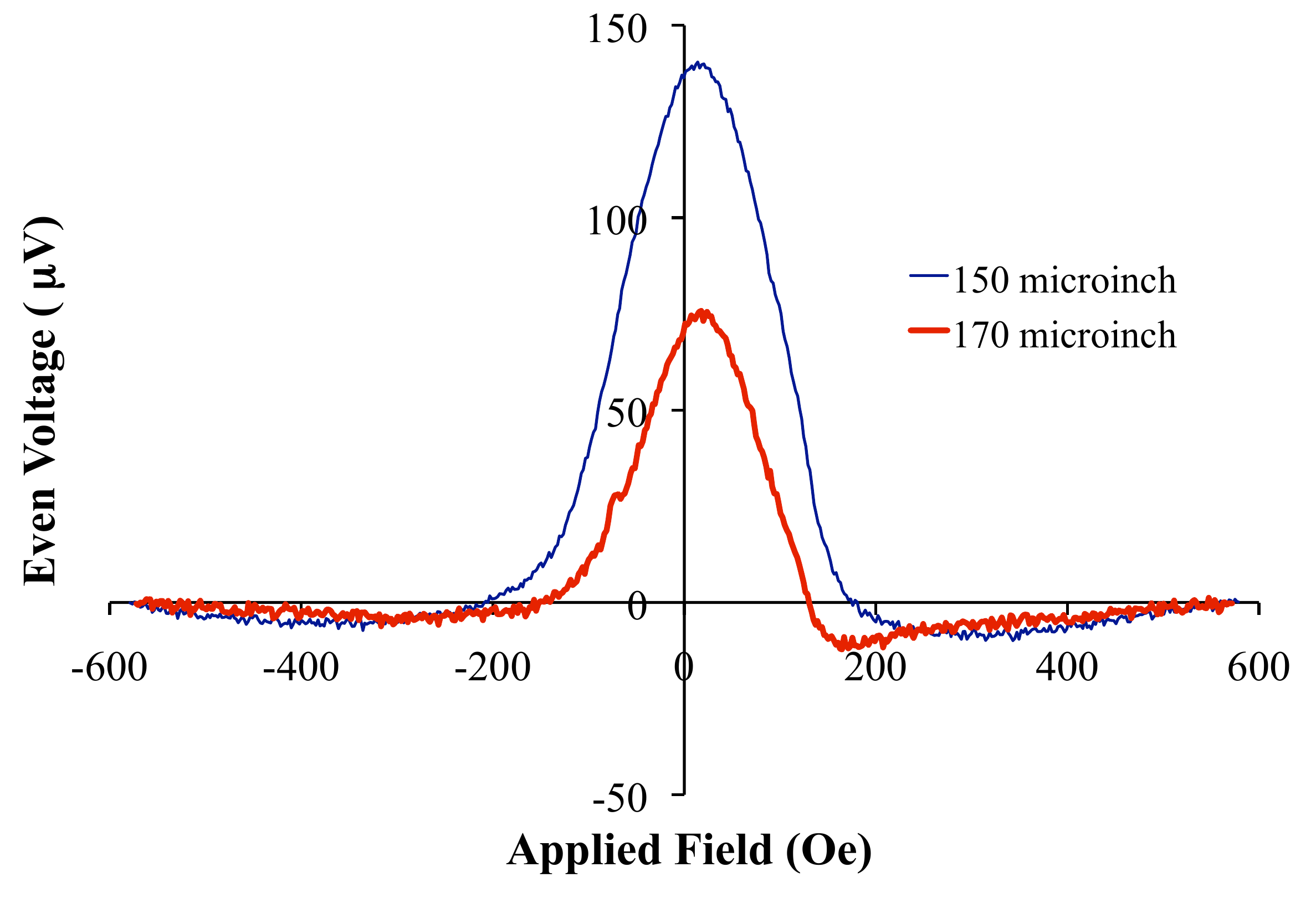}
\caption{The $M$-even voltage observed on YIG samples with different surface roughness. The samples were ground with 120 (80) grit abrasive which resulted in an average surface roughness $R_a$ of 150 (170) microinches. The surface roughness was measured using a Mitutoyo Surftest -- 402 surface roughness tester. In both cases copper electrodes were bonded to the YIG surfaces using silver epoxy.\label{fig10}}
\end{figure}

We also studied the effect as a function of the roughness of the YIG surfaces facing the electrode and discovered that the effect diminishes in size as the surfaces become rougher, at least up to a point (Fig.~\ref{fig10}).\footnote{This observation may partially explain why the $M$-even effect was reduced after the edm toroid was reassembled. In the 2005 experiment the electrode surfaces of the garnet were polished flat by the manufacturer while in our most recent measurements we ground them with 600 grit abrasive, creating a rougher surface.} We speculate that in the process of roughening the surface, certain crystal orientations may be more likely to be removed from the surface. The remaining, more durable crystal orientations might then dominate the surface magneto-electric and magneto-mechanical effects. Alternatively, it may simply be that the size of the effect is proportional to the fraction of the surface that is in direct contact with the electrode. However, no clear trends emerged in the data when the metallic electrodes were similarly roughened. The $M$-even voltage also shows no dependence on the thickness of the electrode for thicknesses between 0.05 mm and 0.5 mm. 

We found significantly different $M$-even responses in more recently fabricated YIG samples. The manufacturer investigated and found that the newer samples had grain sizes of about 12 microns while the older samples had grain sizes of about 8 microns.\cite{fleming} We chose to use only the older samples with smaller grain size in this investigation. Most of our subsequent $M$-even observations have been standardized to relatively rough surfaces ($R_a \approx 170\ \mathrm{microinches}$) and relatively high pressure (about 490 psi).

\begin{figure}
\includegraphics[scale=\kscale]{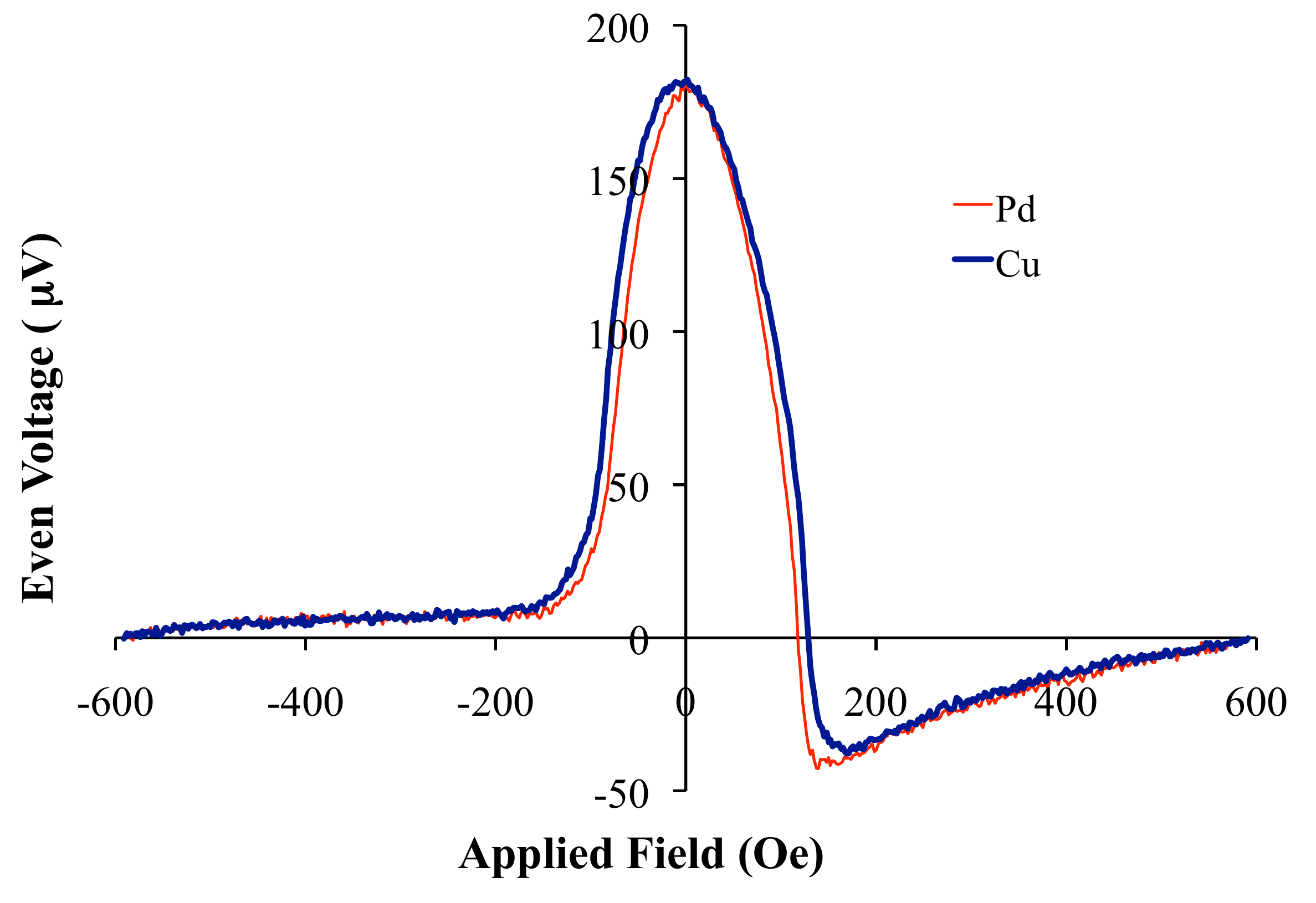}
\caption{Palladium (0.1 mm, 99.9\%) and copper (0.05 mm, 99.98\%) $M$-even curves.\label{fig11}}
\end{figure}

\begin{figure}
\includegraphics[scale=\kscale]{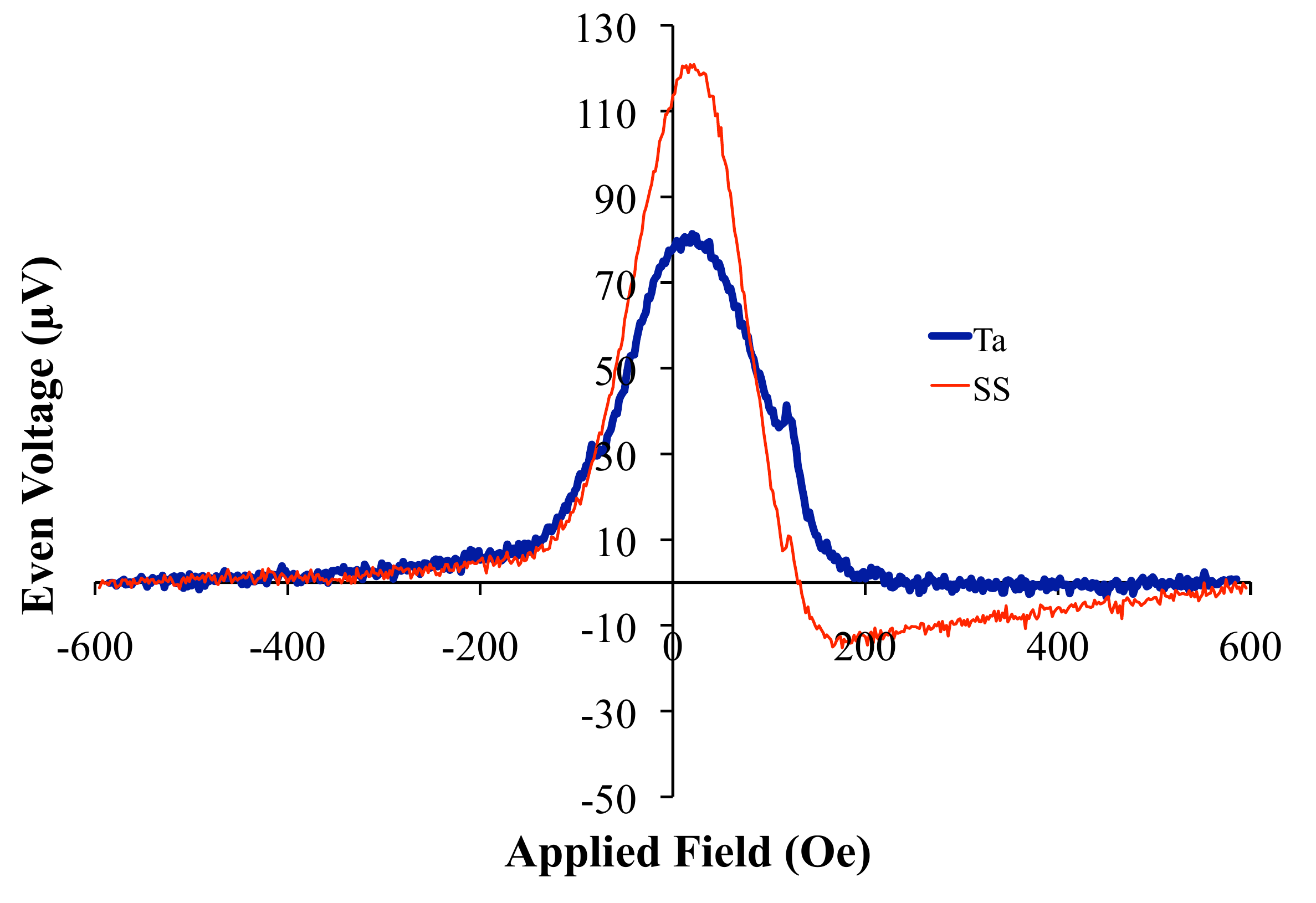}
\caption{Stainless steel (0.05 mm) (fine) and tantalum (0.127 mm, 99.9\%) (bold) $M$-even curves.\label{fig12}}
\end{figure}

\begin{figure}
\includegraphics[scale=\kscale]{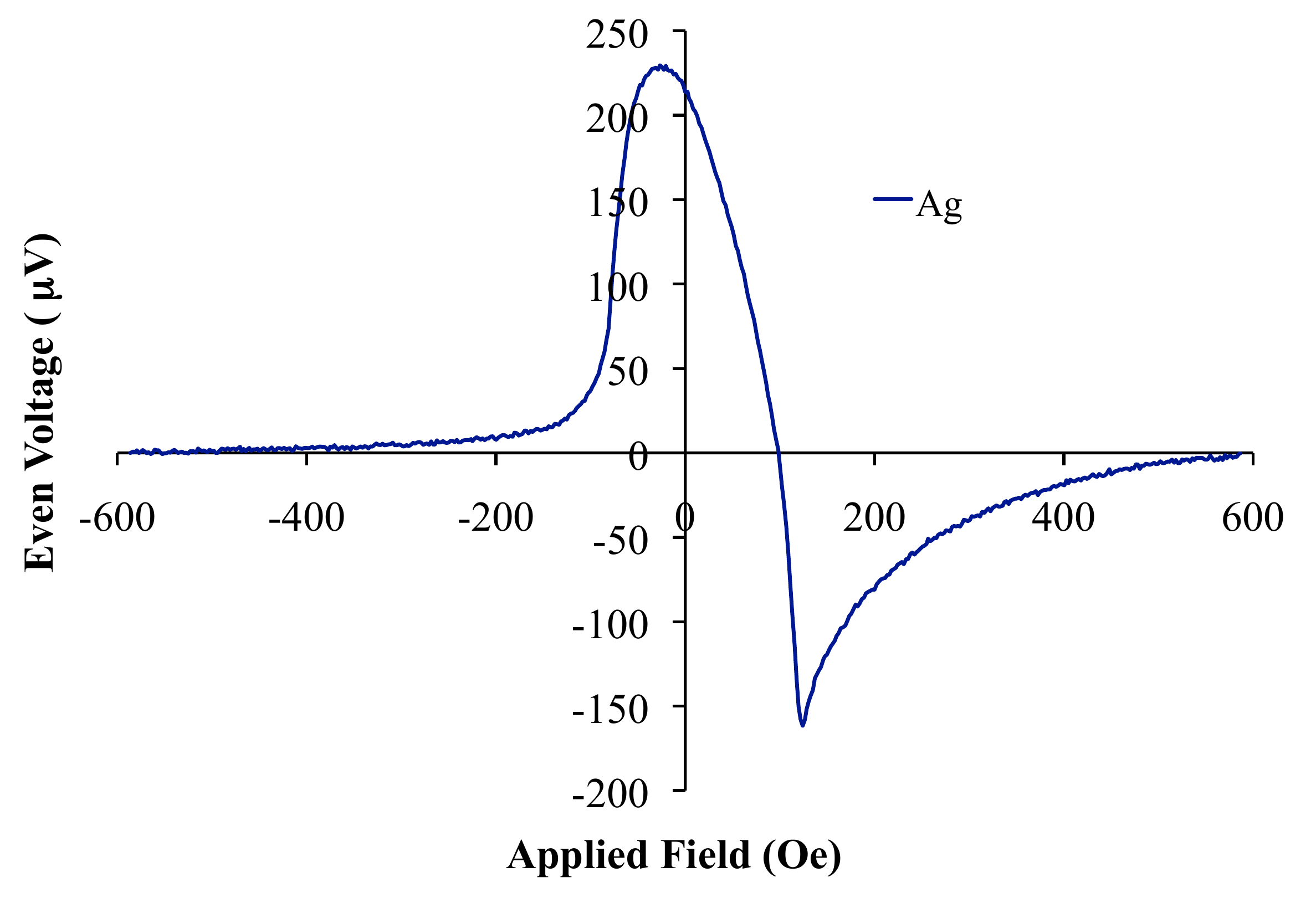}
\caption{The $M$-even voltage observed with silver (0.05 mm, 99.9\%).\label{fig13}}
\end{figure}

\begin{figure}
\includegraphics[scale=\kscale]{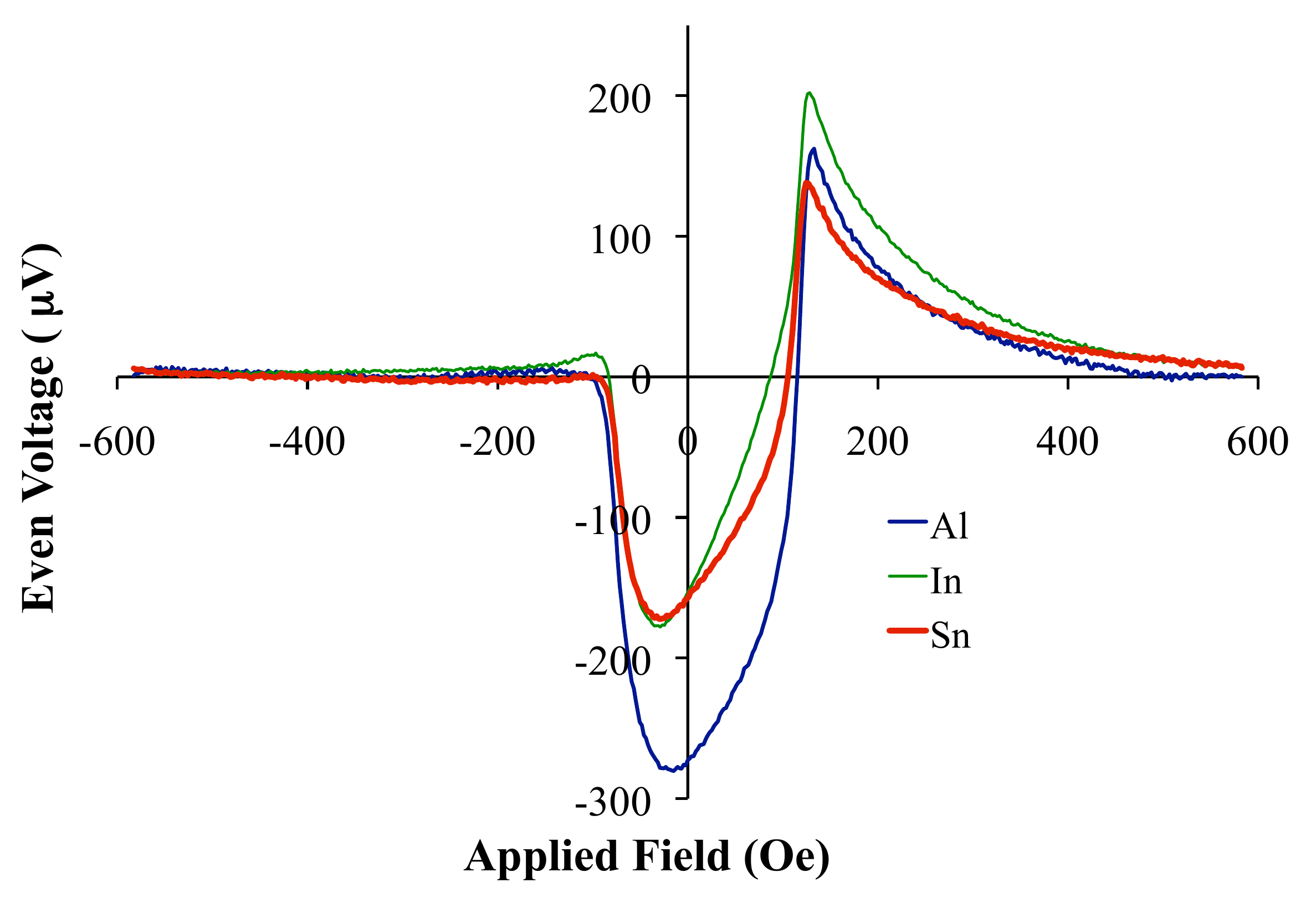}
\caption{The $M$-even voltages observed for aluminum (0.1 mm, 99.99\%), indium (0.127 mm, 99.99\%) and tin (0.05 mm, 98.8\%).\label{fig14}}
\end{figure}

We measure the voltages produced on a variety of metallic electrode materials and find a wide range of different $M$-even responses. Copper and palladium yield $M$-even curves that are virtually identical (Fig.~\ref{fig11}). Stainless steel and tantalum produce curves of similar shape but different amplitude (Fig.~\ref{fig12}). They both exhibit a characteristic ``blip'' as the magnetization approaches saturation. Silver has a large central maximum, but exhibits an abrupt change of sign as the magnitude of $M$ is increased, followed by a large negative tail as the sample approaches saturation (Fig.~\ref{fig13}). The ``poor metals'' (aluminum, indium and tin) produce $M$-even voltages that have the same shape as silver, but with the opposite sign (Fig.~\ref{fig14}). 

\begin{figure}
\includegraphics[scale=\kscale]{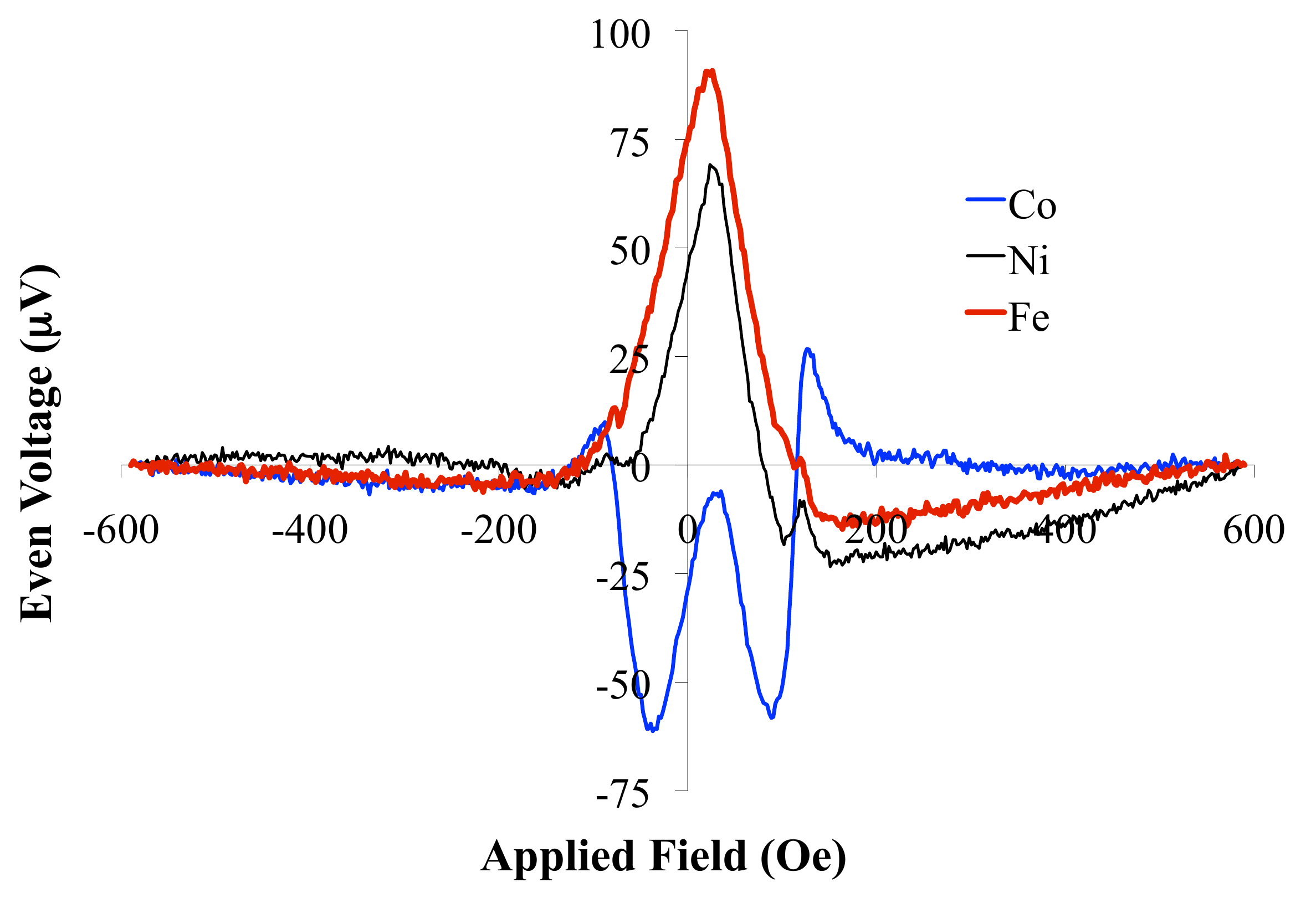}
\caption{The $M$-even response of iron (0.1 mm, 99.99\%), nickel (0.1 mm, 99.5\%) and cobalt (0.1mm, 99.95\%).\label{fig15}}
\end{figure}

Ferromagnetic electrodes (Fe, Ni and Co) tend to produce $M$-even voltages that have sharper low field peaks and more modest amplitudes, especially at higher applied fields (Fig.~\ref{fig15}). Cobalt clearly displays additional structure that is only suggested in the other plots. 

\begin{figure}
\includegraphics[scale= \kscale]{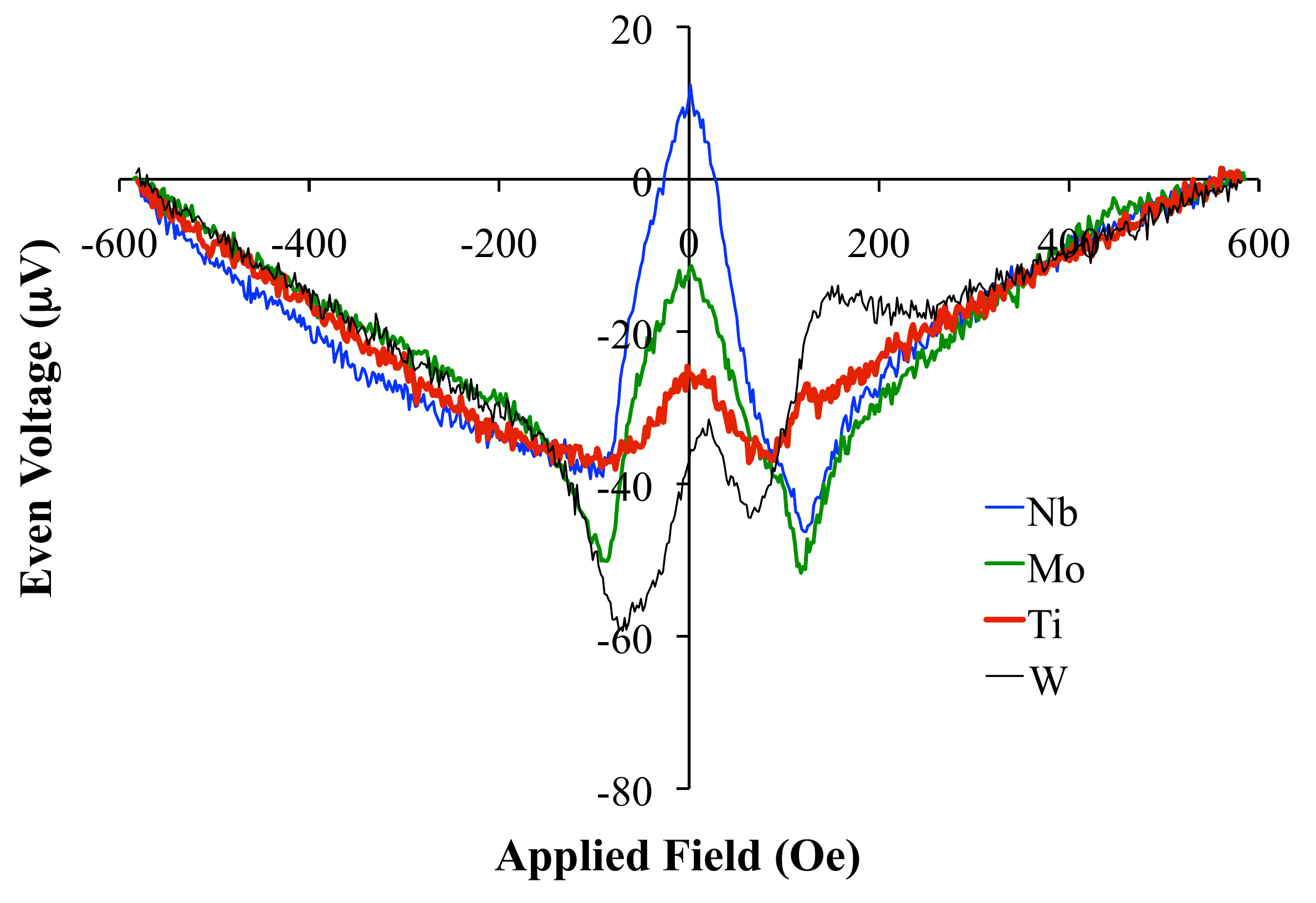}
\caption{The $M$-even response of niobium (0.05 mm, 99.8\%), molybdenum (0.1 mm, 99.5\%), titanium (0.127 mm, 99\%) and tungsten (0.05 mm, 99.95\%).\label{fig16}}
\end{figure}

Other hard non-magnetic electrodes (Nb, Mo, Ti and W) produce small low-field peaks and rather complex shapes similar to cobalt (Fig.~\ref{fig16}). However, the voltage amplitude tends to decay slowly as the applied field is increased. The high-field $M$-even signals for these metals are strikingly similar.

\begin{figure}
\includegraphics[scale=\kscale]{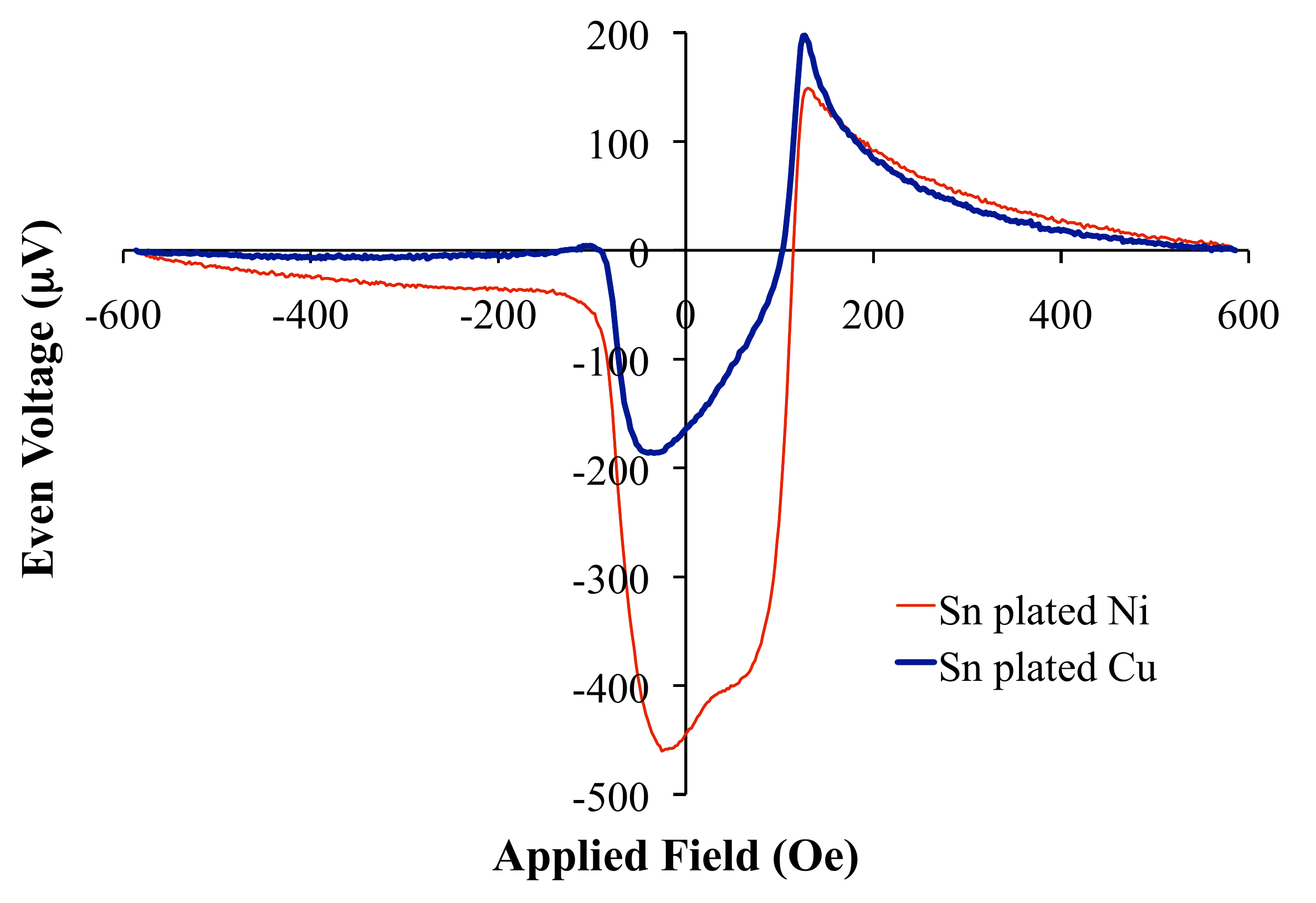}
\caption{The observed $M$-even voltage for copper and nickel electrodes plated with tin.\label{fig17}}
\end{figure}

Further evidence that the effect is primarily produced at the garnet-metal interface is provided by measurements with plated electrodes. Fig.~\ref{fig17} shows the $M$-even effect observed on tin-plated nickel and tin-plated copper. While there is significant variation in the amplitude of the signals, the sign and general shape of the signal is similar to that of tin and is clearly different from that obtained from both copper and nickel.

\begin{figure}
\includegraphics[scale=\kscale]{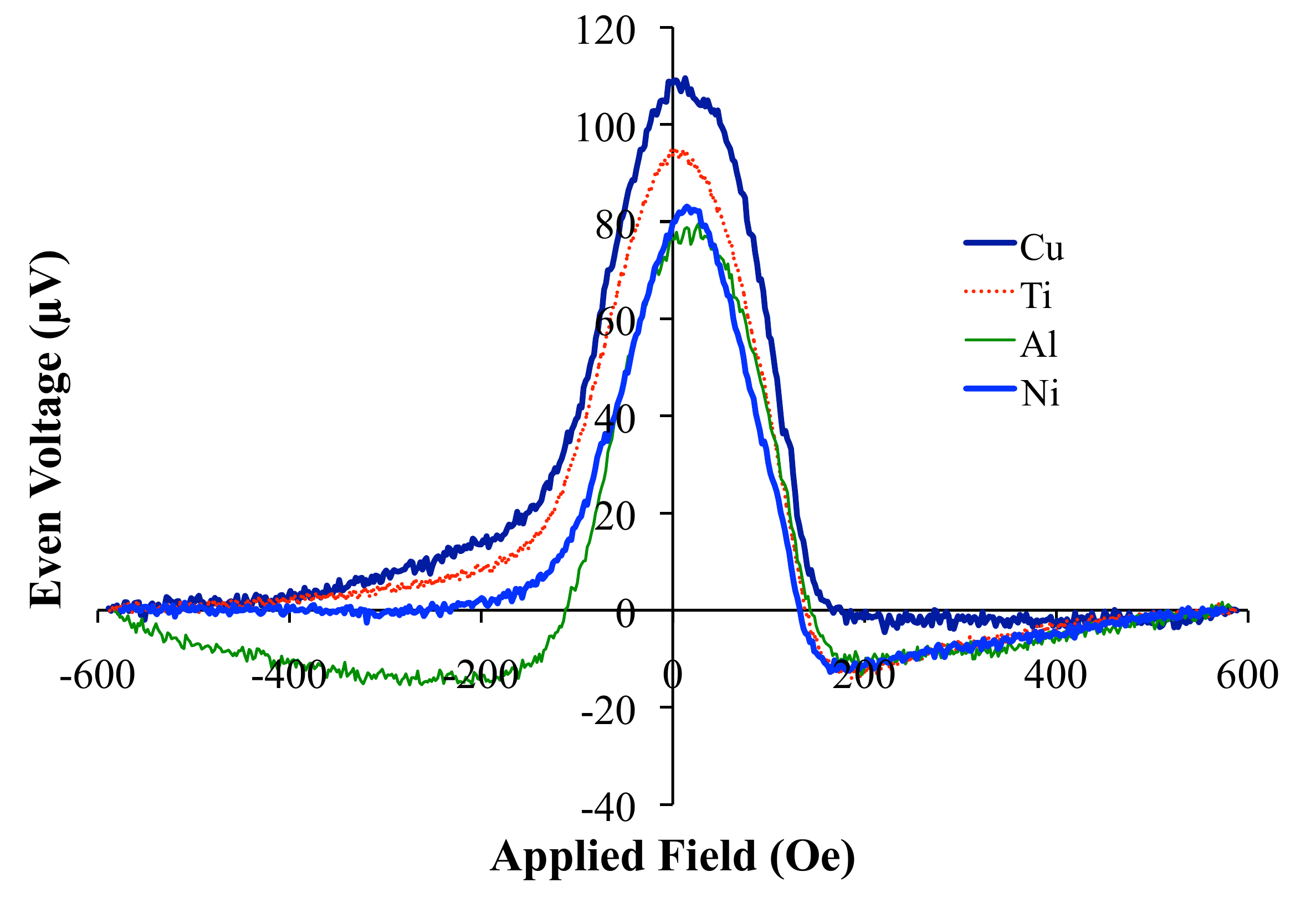}
\caption{The $M$-even signal from four different electrodes bonded with silver impregnated epoxy.\label{fig18}}
\end{figure}

We have studied the effect of bonding various metals to the YIG using different epoxies. Figure \ref{fig18} shows the $M$-even signals associated with four different metals bonded with silver-impregnated epoxy (Epo-Tek EE129-4). The 4 curves are remarkably similar, providing further evidence that the effect is dominated by the surface in immediate contact with the YIG.

The epoxy composition can play an important role in the observed $M$-even signal. The difference between nickel electrodes bonded with nickel epoxy (Epo-Tek N20E) and with non-conductive epoxy (Epo-Tek 377) is shown in figure 19. It is interesting to note that the nickel impregnated epoxy is more sharply peaked at the center, similar to the results obtained from the unbonded magnetic electrodes (Fig.~\ref{fig15}). 

\begin{figure}
\includegraphics[scale=\kscale]{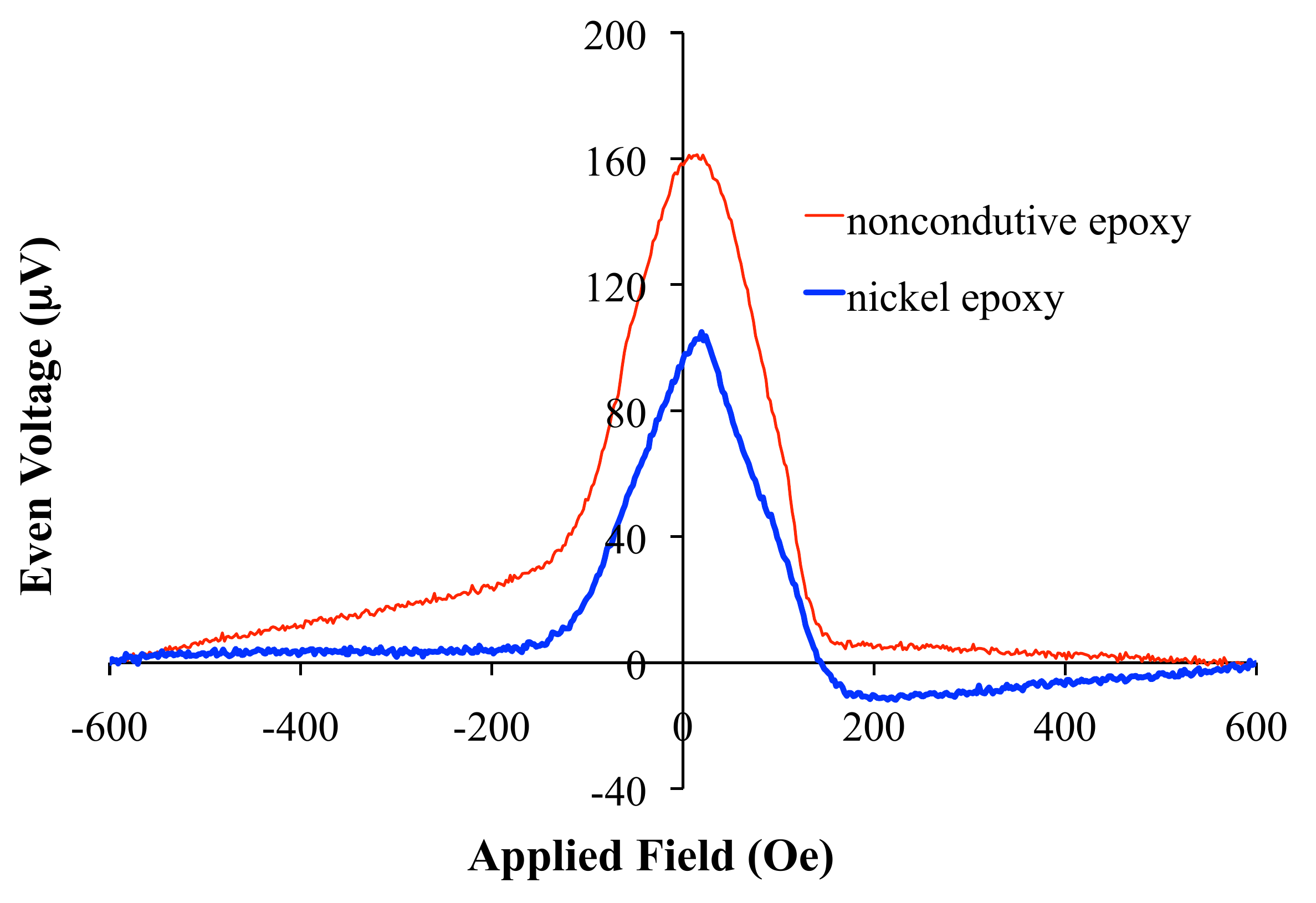}
\caption{The $M$-even effect observed for nickel electrodes bonded with either a nonconductive epoxy or a nickel-impregnated epoxy.\label{fig19}}
\end{figure}

\begin{figure}
\includegraphics[scale=\kscale]{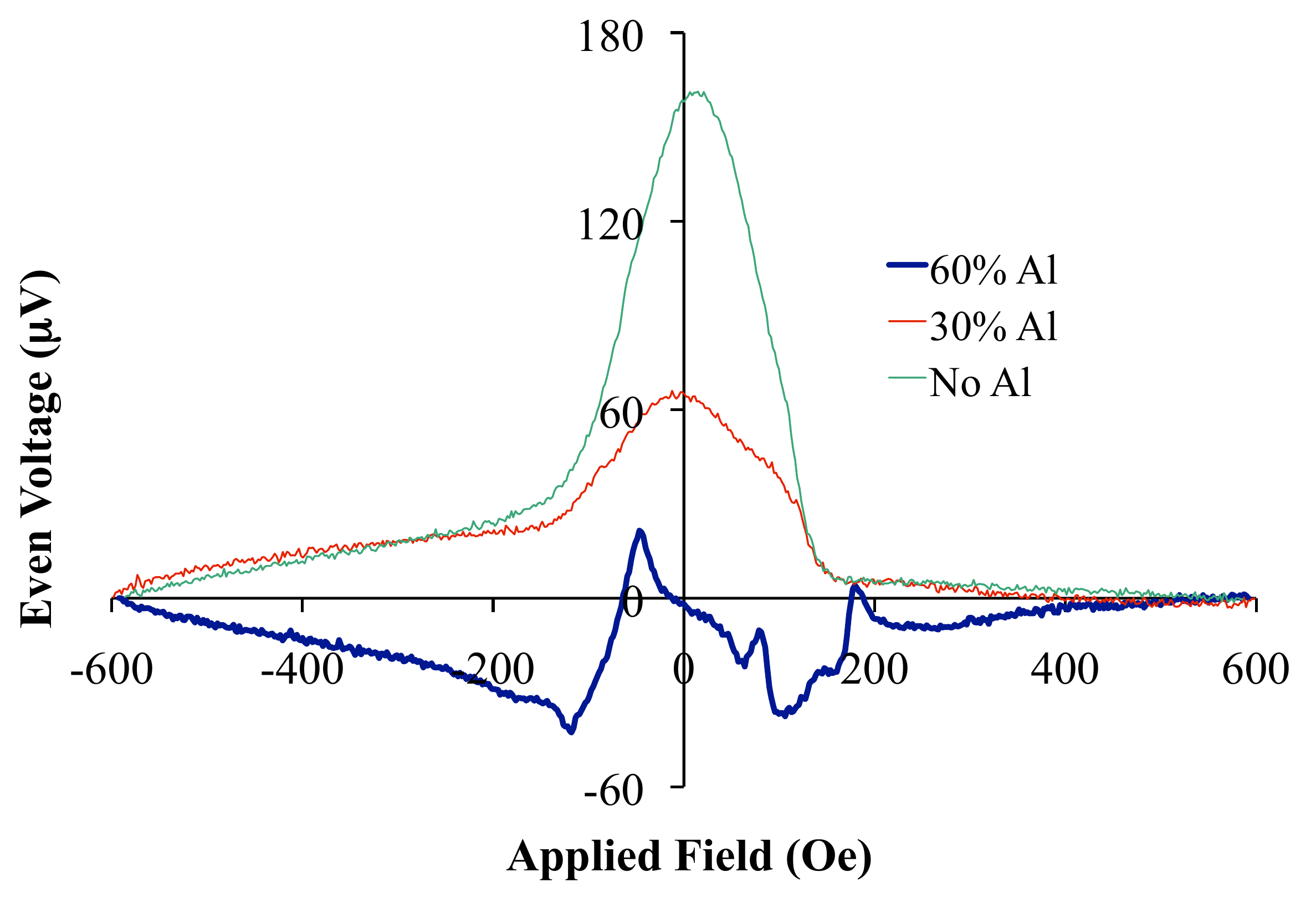}
\caption{The $M$-even signal obtained from nickel electrodes bonded with non-conductive epoxy impregnated with aluminum flake. The fraction is the relative mass of the added aluminum compared with the epoxy.\label{fig20}}
\end{figure}

We note that the signal observed for unbonded aluminum (Fig.~\ref{fig14}) has the opposite sign to that observed with nonconductive epoxy (Fig.~\ref{fig19}). In the hope of minimizing the $M$-even effect, we mixed various fractions of aluminum flake into the non-conductive epoxy and recorded the associated $M$-even voltages. The results are shown in Figure \ref{fig20}. Note that at about 60\% aluminum by weight that the resulting signal achieves a minimum size. The cancellation is, however, not complete and results in a rather complex magnetic dependence. 

\section{Phenomenology}

Based upon these studies we conclude that the $M$-even effect is a magnetically dependent potential generated at the interface between the garnet and the surface in immediate contact with it. Unfortunately, we have not yet been successful at formulating a theoretical model that can account for all of the rich behavior observed. However, we have identified an interesting pattern in the data. All of the curves can be approximately reproduced by superposing two distinct signals. The first is a potential that is quadratic in the observed magnetization (Fig.~\ref{fig8}) evaluated at a field shifted by $H_{\mathrm{shift}}$ with respect to the applied field $H$:

\begin{equation}
V=A\left[M^2 (H-H_{\mathrm{shift}})-M_{\mathrm{sat}}^2\right]
\end{equation}

\begin{figure}
\includegraphics[scale= \kscale]{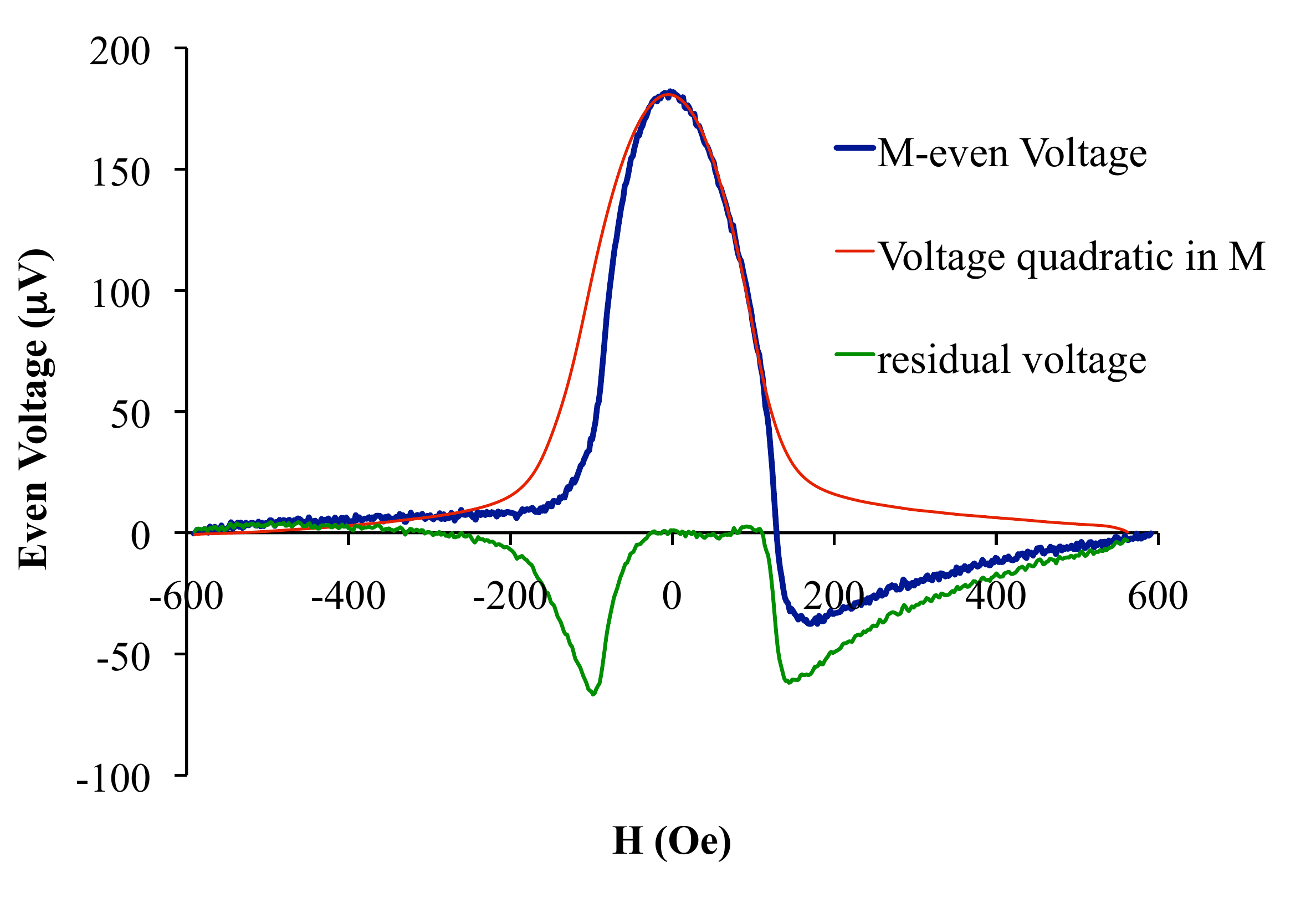}
\caption{The decomposition of the observed $M$-even voltage into a part that is quadratic in $M$ and a residual. This particular sample was an unbonded copper electrode. The quadratic part has been shifted by 28 Oe to the left in order to achieve the fit ($H_{\mathrm{shift}}=-28\ \mathrm{Oe}$). Notice that the central part of the curve is well accounted for and that the residual ``wings'' appear to have an abrupt onset at about 100 Oe.\label{fig21}}
\end{figure}

Here $A$ is a constant and the term $M_{\mathrm{sat}}$ references the potential to zero at saturation. $H_{\mathrm{shift}}$ is a fit parameter that displaces the center of the quadratic distribution. This quadratic term generally accounts for the large and smooth central features of the observed curves. Such an effect may be linked to the magnetostriction of the sample, or more generically to the presence of an $E^2 B^2$ term in the free energy, as discussed below. Once this effect is removed from the data, two symmetric ``wings'' remain (Fig.~\ref{fig21}). These have a sharp onset at a particular value of the applied field and then decay away at higher applied field. This part of the potential may be associated with the surface of the garnet transitioning into modes II and III of the magnetization (or demagnetization) process~\cite{mercier:1974ab} where further alignment (or depolarization) of the magnetization requires rotating the local moment away from (or toward) the easy axis of the crystal. 

Similar decompositions can largely accommodate all of our data. Pd, Ag and stainless steel result in fit parameters with the same sign as copper. Ta is nearly entirely accounted for by the voltage quadratic in $M$. The poor metals (In, Al and Sn) have the opposite signs of both contributions. The narrow and sharp features of the magnetic electrodes (Co, Ni and Fe) can be accounted for with relatively small quadratic parts and sizable ``wings'' which have their onset at much smaller applied fields. This suggests that the magnetization of the YIG's surface at an interface with a magnetic electrode requires more modest coercive fields due to the better flux match at the interface.

\section{A Simple Model}

Bulk magnetoelectric effects have been observed in single-crystal YIG~\cite{odell:1967ab} and GdIG.\cite{lee:1970ab} Because these Garnet crystals are centro-symmetric, no linear magnetoelectric effect is possible. Legg and Lanchester have suggested that the bulk magneto-electric effects observed in YIG can be accounted for by electrostriction and inverse magnetostriction.\cite{0022-3719-13-35-014} This effect can be parameterized by introducing a term in the free energy
\begin{equation}
F=\gamma\, E^2 B^2
\end{equation}
where $B$ ($E$) is the total magnetic (electric) field and the tensor indices have been omitted.\cite{PhysRevA.79.022118} Notice that this free energy is fully allowed from symmetry considerations. Differentiation with respect to $E$ yields the electric displacement, $D$:
\begin{equation}
D=\frac{d F}{d E}=2\,\gamma\, E B^2
\end{equation}
Hence, the electric field associated with the magneto-electric effect is 
\begin{equation}
E_{\mathrm{ME}}=\frac{D}{\kappa}=\frac{2\, \gamma\, E B^2}{\kappa}
\end{equation}
where $\kappa\approx15$ is the dielectric constant for YIG. We imagine that the presence of the metal electrode at the YIG surface can induce a contact field $E$ and an associated contact potential $V = E\, d$, where $d$ is the distance that the contact field penetrates into the YIG surface. In turn, the magneto-electric field will induce a magneto-electric voltage at the surface 
\begin{equation}
V_{\mathrm{ME}}=d\, E_{\mathrm{ME}}=\frac{2\, \gamma\, d\, E B^2}{\kappa}=\frac{2\, \gamma\, V B^2}{\kappa}
\end{equation}
Hence, this simple model predicts a magneto-electric voltage at the electrode surface that is linear in the contact potential and quadratic in $B$. Since $B$ is dominated by $M$ in the YIG, this term would qualitatively agree with the part of the $M$-even voltage that appears to be approximately quadratic in $M$. In this model, the observed variation of the magnitude and sign of the $M$-even effect is attributed to different values of the contact potential between the YIG and the adjacent surface. 

\begin{figure}
\includegraphics[scale=\kscale]{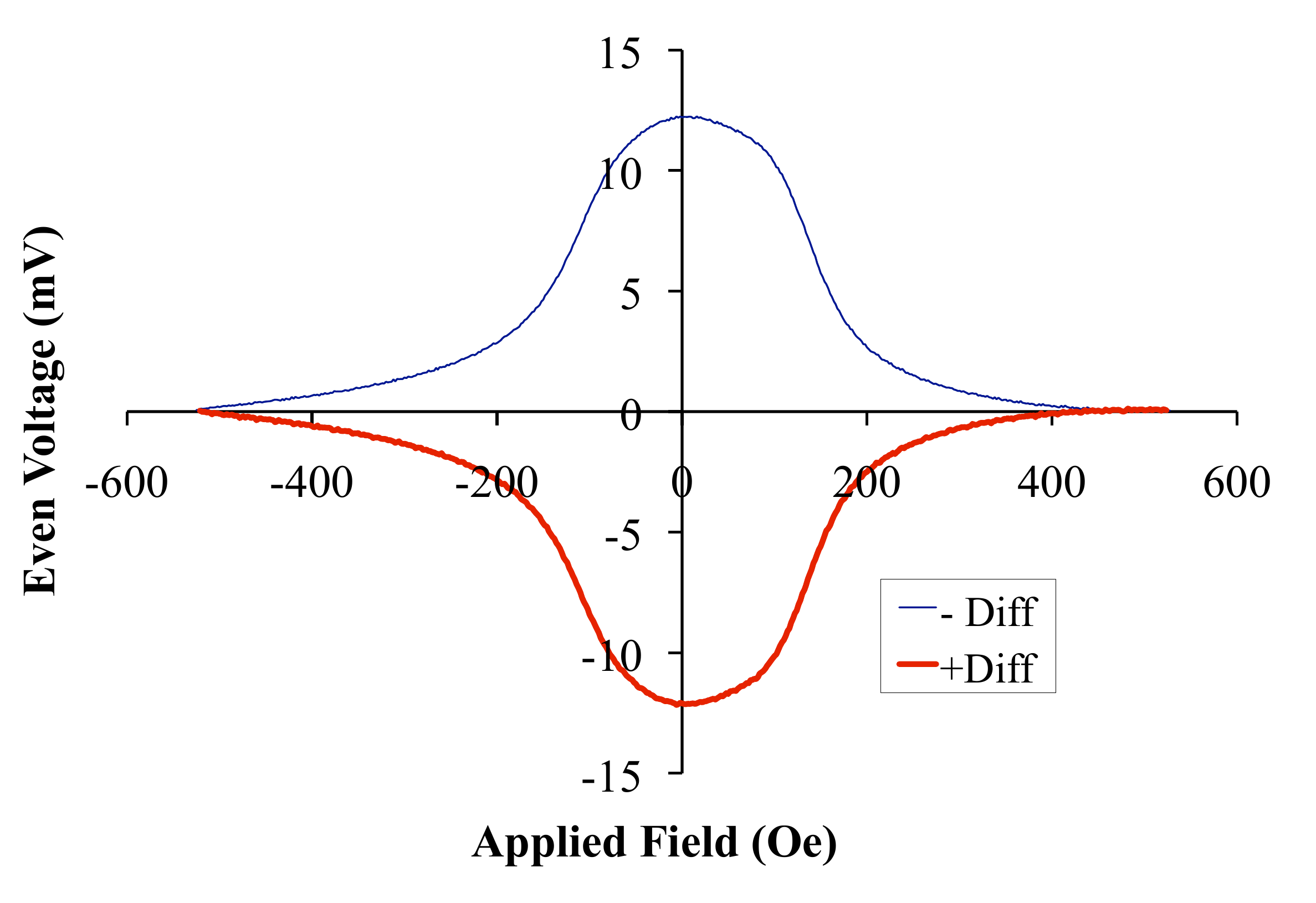}
\caption{The magnetoelectric voltage even in the applied field observed on a polycrystaline YIG sample. The curve $+\mathrm{Diff}$ ($-\mathrm{Diff}$) is the difference between two traces, one taken with the center electrode at $+500\ \mathrm{V}$ ($-500\ \mathrm{V}$), and one taken with the center electrode at 0 V. The Faraday cage surrounding the sample and the electrode painted on the outer ends of the YIG pieces are held at earth ground. These smooth signals represent the contribution to the voltage due to the bulk magnetoelectric effect.\label{fig22}}
\end{figure}

\begin{figure}
\includegraphics[scale=\kscale]{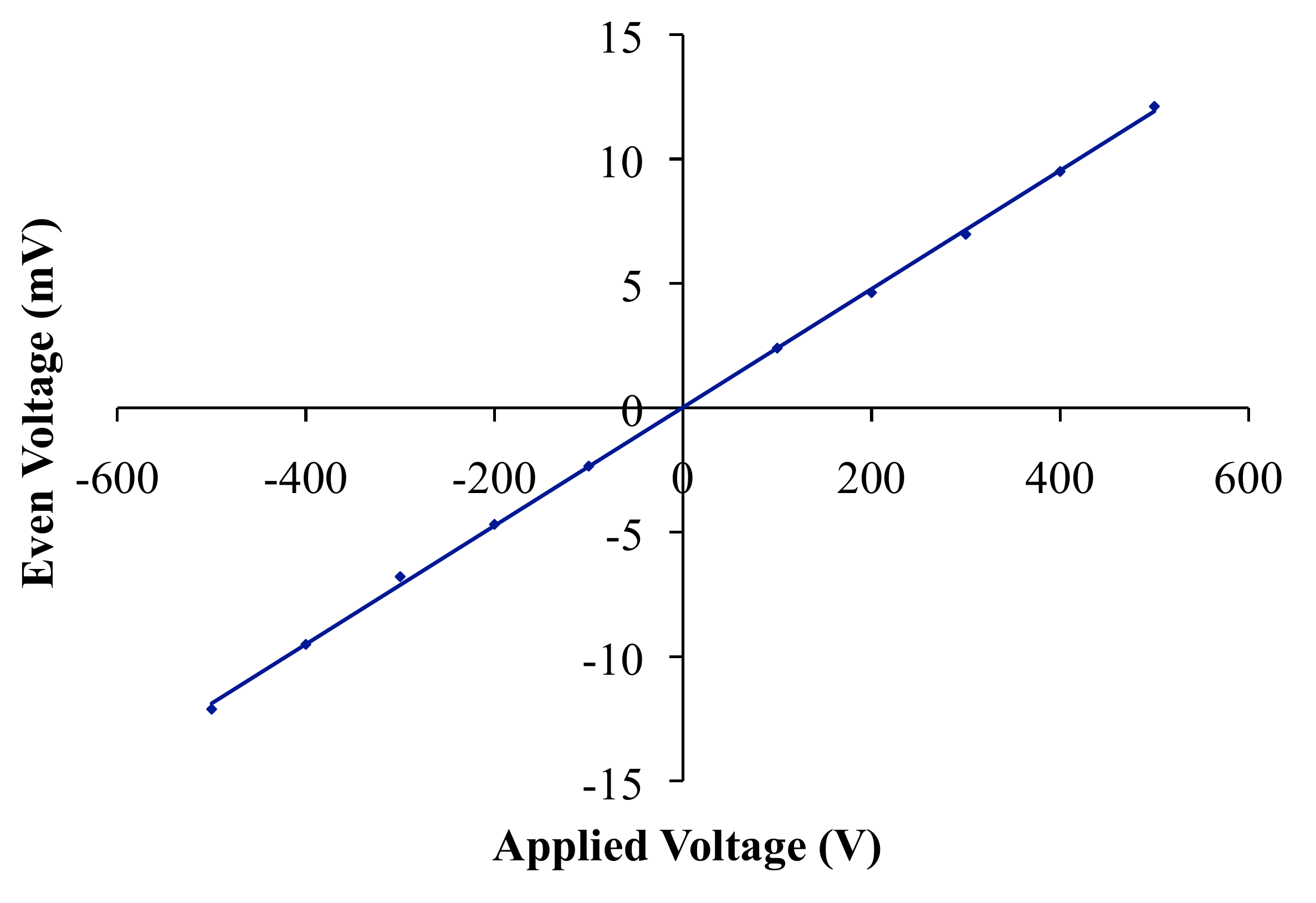}
\caption{The difference between the electrode voltage when the magnetization is saturated and when the magnetization is zero. The slope of the line is $2.38\times10^{-5}$.\label{fig23}}
\end{figure}

To investigate this possibility further we observe the YIG's bulk magneto-electric response. To do this, we apply a voltage to the center electrode while the two exterior ends of the cylinder, which have been coated with conductive silver paint, are held at ground. The signal from the center electrode is AC coupled to our detector using a 47 pF blocking capacitor and is isolated from the supply voltage by a $2\times10^{11}\ \Omega$ resistor. The applied magnetic field is modulated at about 1 Hz using a triangle wave. The $H$-even potentials observed in our polycrystalline samples are shown in Figure \ref{fig22}.

The magneto-electric response of our sample appears to be approximately quadratic in the sample magnetization and linear in the applied electric field (Fig.~\ref{fig23}). The ratio of the maximum magneto-electric voltage change (between the magnetization of zero and saturation) and the applied voltage is $2.4\times10^{-5}$. We estimate the calibration uncertainty in this ratio to be about 10\%. This ratio is a little less than half of the value reported by O'Dell for a single crystal along the 110 axis.\footnote{ Note that O'Dell appears to correct his plotted data by a factor of 3 to accommodate capacitive losses.}

Our bulk magnetoelectric measurements suggest that to produce an effect comparable in size to the quadratic part of our observed $M$-even voltages (typically $100\ \mathrm{\mu V}$) would require a potential drop at the YIG surface of about 4 Volts. This seems a bit large for a surface contact potential but is not completely unreasonable. We conclude that this attractively simple model may be adequate to account for the part of the observed $M$-even effect quadratic in $M$. The $M$-even ``wings'' remain unexplained. Further, we are not able to account for the various magnitudes and signs of the contact potential differences required to match the different electrode materials. Clearly, a more rigorous and detailed model, perhaps one that considers some of the additional mechanisms outlined in ref.~\cite{PhysRevLett.101.137201}, will be required to achieve a quantitative understanding of the $M$-even effect.

\begin{acknowledgments}
We thank Dr.~Daniel Krause, Jr., Norman Page, Robert Bartos and Robert Cann for technical support and Prof. Jonathan Friedman for important discussions. This work was funded by NSF grant PHY-0555715 and Amherst College. LRH would also like to thank Prof. David DeMille, Prof. Steven Lamoreaux and Yale University for their hospitality during part of this work.
\end{acknowledgments}

\bibliography{M_even}

\end{document}